\pdfoutput=1
\documentclass[journal,preprint]{vgtc}

\vgtccategory{Research}

\preprinttext{Preprint}
\manuscriptnote{}

\title{Vega-Video: Integrating Video into the Grammar of Graphics}

\author{%
  \authororcid{Dominik Winecki}{0009-0009-8632-3102} and
  \authororcid{Arnab Nandi}{0000-0002-4138-603X}
}

\authorfooter{
  \item
    Dominik Winecki and Arnab Nandi are with The Ohio State University.
    E-mail: \{winecki.1\,$|$\,nandi.9\}@osu.edu.
}

\abstract{%
Video data is increasingly used alongside conventional data for interactive data exploration, necessitating interfaces for exploring and presenting mixed-modality data.
However, integrating video into visualizations remains difficult due to its distinct paradigms and inherent performance challenges.
We identify three classes of video data visualization -- synchronization, annotation, and transformation -- and integrate them into the Vega declarative grammar.
We show that these abstractions enable high-performance implementation.
To reconcile Vega's instantaneous dataflow with video player state, we introduce a split-signal architecture that preserves declarative semantics while masking video update delays.
We detect continuous scrubbing interactions at compile time to apply encoding-aware optimizations that improve responsiveness by up to $4\times$.
We also repurpose VOD protocols to transform videos in real time, delivering sub-200ms updates even on multi-hour-long compilations.
These contributions enable seamless integration of conventional and video data visualization.
}


\graphicspath{{figs/}{figures/}{pictures/}{images/}{./}}

\usepackage{amsmath}
\usepackage{amssymb}
\usepackage{enumitem}
\usepackage{booktabs}
\usepackage{mathptmx}
\usepackage{flushend}

\begin{document}

\firstsection{Introduction}

\maketitle
\hypersetup{pdfsubject={Preprint}}

We consider interactive visualizations combining \textit{video data} with \textit{conventional---non-video---data}.
Multimodal data pipelines collect significant volumes of associated metadata, tabular data, and time-series data, and generate derived datasets and computer vision model inference outputs to complement video data streams.
Video data is no longer a stand-alone medium; regardless of the application, additional data is captured alongside it or extracted from it.
Just as conventional data requires conventional visualization, hybrid conventional/video data requires multimodal conventional/video visualization.

Creating data visualizations that incorporate video data remains challenging.
Semantically, most modern tools build on the grammar of graphics~\cite{wilkinson_grammar_2005}, which is conceptually incongruent with incorporating video player elements or transforming video data.
Additionally, dataflow graph architectures for dynamic/interactive visualizations assume a consistent, unified state with instantaneous propagation.
However, resynchronizing video playback state is relatively slow, quickly exceeding the strict latency constraints required for a good user experience.
Video systems are inherently difficult due to the volume and computational requirements of videos.
A single frame can contain more information than an entire data dashboard.
Further, videos are encoded/compressed out of necessity, and only a small portion is kept in memory at a time.
These factors are antithetical to the traditional principles of high-performance visualization.
Hardware acceleration and decades of optimizations make high-resolution video playback possible, even on commodity devices.
However, the video playback state is still subject to what is, in essence, \textit{inertia}.
A video in playback will continue to play without delay, but initially opening the video, seeking to a new position, or changing video content, i.e., \textit{interactions}, still occur on the scale of hundreds of milliseconds to multiple seconds.
Such inertial resistance to interaction creates detrimental lag across the entire visualization when a unified approach to visualization/playback state is used.

These challenges arise against a backdrop of rapid growth in video data driven by the proliferation of high-resolution cameras, declining storage and compute costs, and unprecedented advances in machine learning models' visual understanding.
Cameras have become the primary way to observe the physical world --- the ultimate sensor --- making video a quintessential part of multimodal data pipelines.
In many of these applications, humans are uniquely capable of integrating subtle spatial, temporal, and contextual cues in video with structured summaries in charts or tables.
It is therefore critical that they be equipped with useful video data visualization tools that leverage humans' unmatched visual reasoning across both modalities.

Consider a researcher studying vehicular near-misses and aggressive driving events in a fleet equipped with dash-cams and telemetry monitors.
Per-event visualization may involve synchronizing time-series telemetry plots with video feed(s), charting trajectories on a map, annotating the videos with lane markers or object detections (e.g., cars, people, bikes), and displaying event metadata.
Interactions would be playing the video and jumping to interesting positions on the timeline.

Beyond single-event visualization, cross-event visualization covers the full dataset.
Each event corresponds to a tuple, and correlations and aggregations can be plotted; subsequent interactions can filter the events.
While this is a typical exploration approach, adding video-native elements can provide further value.
A video compilation can display all events and be linked to histograms or scatter plots, with brushing triggering real-time filtering of the video compilation.
Further, compilations can be sorted by an attribute, such as event time-of-day, so users can use the video timeline progress bar to quickly visualize how different percentiles of the attribute appear in the videos.
For example, the start of the compilation corresponds to early-morning events, the middle to midday events, and the end to evening events.
This linking can also be highly integrated and bi-directional.
A scatter plot, for example, can highlight the currently displayed video event, and clicking on a point in the scatter plot can jump the video to that event.
Visualizations that can edit and transform videos --- beyond just syncing and annotating them --- are clearly powerful exploration paradigms, but are either impractical or impossible under current systems.

The preceding motivating example spans from long-established techniques to novel ones.
We split these into three categories: syncing, annotating, and linked video transformation.
Syncing with video is well established, with numerous visualization systems supporting it, yet they lack expressivity and portability; many offer only temporal synchronization with line plots, with limited potential for further development, interface exploration, or performance optimization.
Conversely, widely used visualization grammars excel at composing a wide range of graphics linked with interactions.
Annotation is conceptually similar and is relegated to domain-specific tasks, particularly drawing bounding boxes or labels.
Entirely unexplored is linked video transformation, in which interactions with a conventional part of a visualization update the content of a video player, not just its playback state.
This presents a challenge both in the semantics of representing these visualizations and in significant performance hurdles to ensuring low-latency real-time updates.
The closest approach in use is to provide links or embedded players for individual source video files, a primitive approach comparable to displaying a text field.
Few visualization systems support either of the first two video data visualization classes; none support the latter, and none tie these ideas together.

In this paper, we build an extension to Vega~\cite{2014-reactive-vega} to support first-class video data visualization.
We extend its grammar and that of higher-level systems, such as Vega-Lite~\cite{2017-vega-lite}, to support declarative integration of video player elements.
We provide a unified representation of time, making specifications indistinguishable from the existing dataflow paradigm while masking the underlying synchronization control loop and providing semantically correct behavior.
Further, the slowest synchronization-based updates are those that seek videos, so we build on our time handling to classify them and select optimal seeking strategies, reducing user latency during interactions that trigger seeking.
We introduce a video transformation layer that supports real-time video updates via declarative data transformations, performing lightweight editing.
We release our implementation, vega-video.js, as open source software.

\subsection*{Contributions}

In this paper, we extend the Vega grammar to support first-class video data visualization and resolve the performance challenges necessary for interactive use.
Specifically, we contribute:

\begin{itemize}
    \item A declarative grammar that extends Vega to support video players, frame-level annotations, and video transformations as composable primitives, without modifying Vega's core (\Cref{sec:grammar}).
    \item A split-signal synchronization architecture that reconciles Vega's instantaneous dataflow with the inherent latency of video subsystems, preserving Vega's declarative semantics (\Cref{sec:time}).
    \item Compile-time static analysis to classify seek interactions and apply keyframe-aware seeking strategies to reduce scrubbing latency (\Cref{sec:scrubbing}).
    \item Video transformation via VOD manifest rewrites, repurposing existing streaming infrastructure for real-time data-driven editing (\Cref{sec:virtvideo}).
\end{itemize}

\section{Related Work}

\subsection{Data Visualization}
Foundational to modern visualization, Wilkinson's Grammar of Graphics~\cite{wilkinson_grammar_2005} formalized a semantic approach to representing graphics.
D3~\cite{10.1109/TVCG.2011.185} brought efficient data-driven visualization rendering to the web via direct DOM manipulation.
Vega~\cite{2014-reactive-vega} builds on these by providing a JSON-based declarative grammar for interactive visualizations that incorporates events and signals, and uses a dataflow graph for efficient updates.
Vega-Lite~\cite{2017-vega-lite} further builds on Vega by proposing a higher-level and more concise grammar, enabling practical use of more complex visualization paradigms.
Animated Vega-Lite~\cite{9914804} extends Vega-Lite with time as an encoding channel, unifying animation with interactive graphics.
It models time as timer ticks that propagate instantaneously through the dataflow graph.
We emulate this unified abstraction for video, using split signals to mask the latency of video player updates from the rest of the visualization, and using frame time as an encoding channel for annotations.
Recent work has decoupled data processing from visualization specification: VegaFusion~\cite{2208.06631} and VegaPlus~\cite{2024-vega-plus} offload Vega data transforms to server-side databases, while Mosaic~\cite{10297587} routes declarative queries from visualization clients to DuckDB~\cite{10.1145/3299869.3320212}.
DIEL~\cite{9552893} provides a framework for addressing asynchronous states from external data processing in visualizations.
We build on these ideas for video subsystems, where the external system---a browser video player---introduces latency and state management challenges.
However, unlike offloading queries to a database, the video player ``external system'' primarily functions as a linked/coordinated view.
Existing coordinated view research~\cite{9417674} does not consider components with inherent update latency.

\subsection{Video Data Visualization}
Video has long been recognized as a specialized datatype complementary to conventional data.
In seminal work, Mackay \& Davenport stated that ``video becomes an information stream, a data type that can be tagged and edited, analyzed and annotated''~\cite{mackay_virtual_1989}.
Joining videos and time-series data is the most developed of these areas, as seen in ChronoViz~\cite{fouse_chronoviz_2011}, a tool for visualizing time-coded information.
Fouse further studied particular interaction techniques for time-series video~\cite{Fouse2013}, including frequent scrubbing, which we address in \Cref{sec:scrubbing}.
Domain-specific techniques have been developed for time-series video visualization, such as BEDA~\cite{kim2013beda} and MUVTIME~\cite{ivapp16}, focusing on physiology and behavioral sciences, respectively.
A particular area of focus has been on navigating large video datasets, often through timeline scrubbing~\cite{Fouse2013,10.1145/3025453.3025821}.

Annotation for data \textit{visualization}, such as drawing inference results onto videos, is ubiquitous; modern CV models use OpenCV~\cite{opencv_library} or Supervision~\cite{Roboflow_Supervision} for this.
Annotation has converged to a few consistent patterns; annotations tend to be either global, such as attention heatmaps or text drawn in a corner, or object-centric.
Labeling videos, as in humans applying labels to frames, often also called ``annotating videos'', is similarly a mainstay technique backing computer vision pipelines.
Many tools exist for this task~\cite{sekachev_opencvcvat_2020,dutta2019vgg,dutta2016via,shrestha2023feva}, primarily using bounding boxes on frames or spans over a timeline.

Video transformation, compilation, and editing have also been explored, but significantly less often.
In early work, DIVA~\cite{mackay_diva_1998} proposes a ``stream algebra'' for annotating and rearranging video stream segments.
This enables data-driven compilations, or \textit{supercuts}~\cite{andersen_super_2017}, but not in interactive settings.
V2V~\cite{winecki2024_v2v} is a declarative video editor for editing and annotating videos through a DSL with database-style optimizations.
Video Lens~\cite{10.1145/2642918.2647366} supports brushing-and-linking style filtering of discrete video events, showing the corresponding event video clip.

Unlike these systems, our focus is on creating general-purpose support for video data visualizations \textit{across} these video and conventional data paradigms.

\section{A Declarative Abstraction for Video Data Visualization}\label{sec:grammar}

Our goal is to represent hybrid conventional/video data visualizations in declarative visualization grammars.
We extend the Vega~\cite{2014-reactive-vega} grammar to support the three interaction classes: syncing, annotating, and linked video transformation.
Our grammar remains faithful to Vega's grammar and the Grammar of Graphics, with the pragmatic goal of enabling users to integrate videos into existing workflows with minimal effort.

\subsection{Video Players \& Signals}

\begin{figure*}[t]
  \centering
  \includegraphics[width=\linewidth]{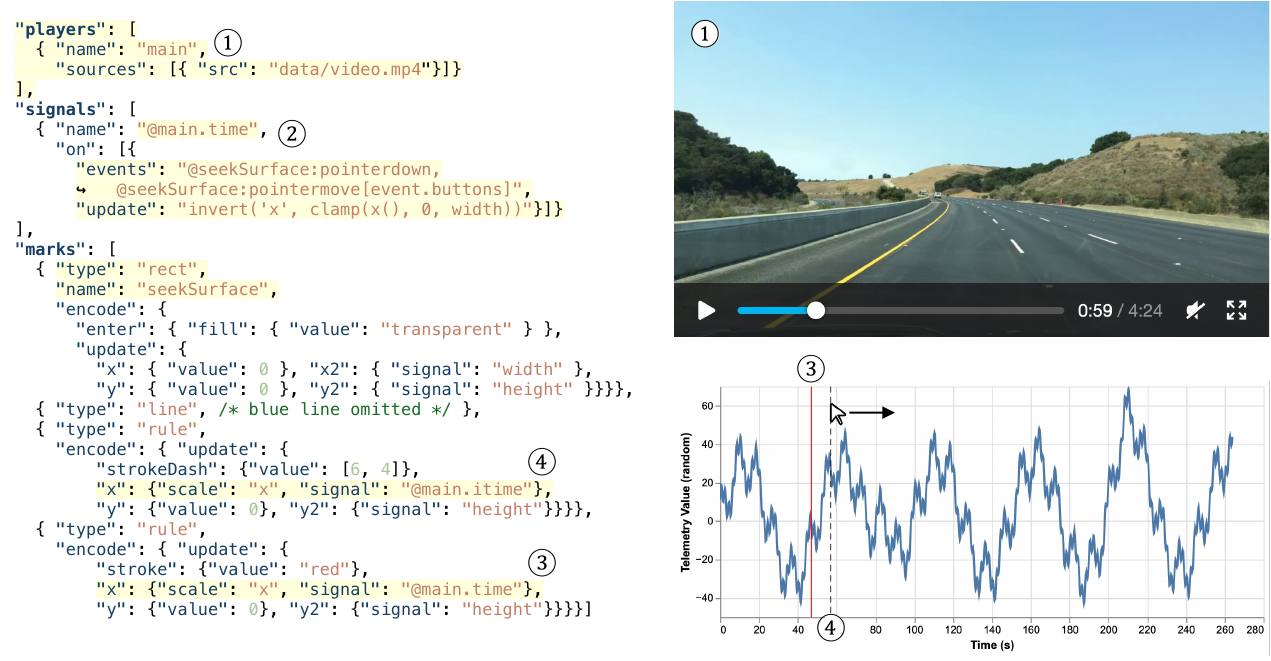}
  \caption{A partial Vega visualization syncing time-series data with a video (\textcircled{1}). Clicking or dragging on the plot updates the video's time (\textcircled{2}). A red playback cursor, at 47s, displays the time of the currently presented frame (\textcircled{3}). A dashed playback cursor at 59s displays the frame the user is targeting (\textcircled{4}).}
  \label{fig:syncing-code}
\end{figure*}

We introduce video players as named, addressable objects in a top-level \texttt{"players"} attribute within a specification.
An example is shown at \textcircled{1} in \Cref{fig:syncing-code}\footnote{Video frames in \Cref{fig:syncing-code,fig:annotation-code,fig:brush_and_link_video,fig:bandl-code,fig:syncing-arch,fig:vod-updates} from the BDD100K dataset~\cite{yu2020bdd100k}.}.
Each player element is configured with a source, either static video or a transformed one.
A player can be positioned by superimposing it on the visualization or by attaching it to an \textit{external} video player in the same web browser.
Video player state is exposed via signals, summarized in \Cref{tab:vega-signals}.

\begin{table}[h!]
    \centering
    \caption{Vega-Video Player Signals}
    \label{tab:vega-signals}
    \begin{tabular}{llc}
        \toprule
        \textbf{Signal} & \textbf{Type} & \textbf{Writable} \\
        \midrule
        \texttt{@player.time} & Float & \checkmark \\
        \texttt{@player.itime} & Float & \checkmark \\
        \texttt{@player.playing} & Boolean & \checkmark \\
        \texttt{@player.duration} & Float &  \\
        \texttt{@player.ready} & Boolean &  \\
        \texttt{@player.ended} & Boolean &  \\
        \bottomrule
    \end{tabular}
\end{table}

The two most important signals are \texttt{@player.time} and \texttt{@player.itime}.
Both represent positions in seconds on the player's timeline, but they capture different aspects:

\begin{itemize}
    \item \textbf{\texttt{@player.time}} is the \emph{presented} time:
        the time corresponding to the frame visible on screen.
    \item \textbf{\texttt{@player.itime}} is the \emph{intended} time:
        the time that the visualization is attempting to show, incorporating user interactions such as rapid scrubs or jumps. This corresponds to the time on the player timeline progress bar.
\end{itemize}

During steady-state playback, these signals coincide.
During interactions that require seeking, however, they diverge:
\texttt{@player.itime} jumps immediately to the target time, enabling instantaneous synchronization between charts as well as video timeline progress bars.
\texttt{@player.time} lags until the video decoder can display the requested frame, but maintains frame-exact correctness.
Video editors, such as Adobe Premiere Pro and Apple Final Cut, exclusively use intended time playback cursors due to their speed.
However, since correctness usually matters in data visualization and exact video interaction techniques are nascent, both are provided, allowing users to choose the desired behavior or combine them, as in \Cref{fig:syncing-code}.
Minimizing the difference between these two values is a central goal of this work and is addressed in \Cref{sec:scrubbing}.

A visualization of synchronization using these signals is shown in \Cref{fig:syncing-code}.
A single video player is loaded with a static video source file.
A seek surface (\textcircled{2}) is created so that pointer selections on the plot update the video's position.
Two playback cursors display each of the ``current'' times, one for the displayed frame (\textcircled{3}) and one for the intended time (\textcircled{4}) during an active interaction.
The visualization state is shown either during a rapid \texttt{pointermove} drag across the plot or during a rapid scrub on the video player's timeline progress bar; otherwise, the two cursors coincide.

\subsection{Video Annotation}

\begin{figure*}
    \centering
    \includegraphics[width=1.0\linewidth]{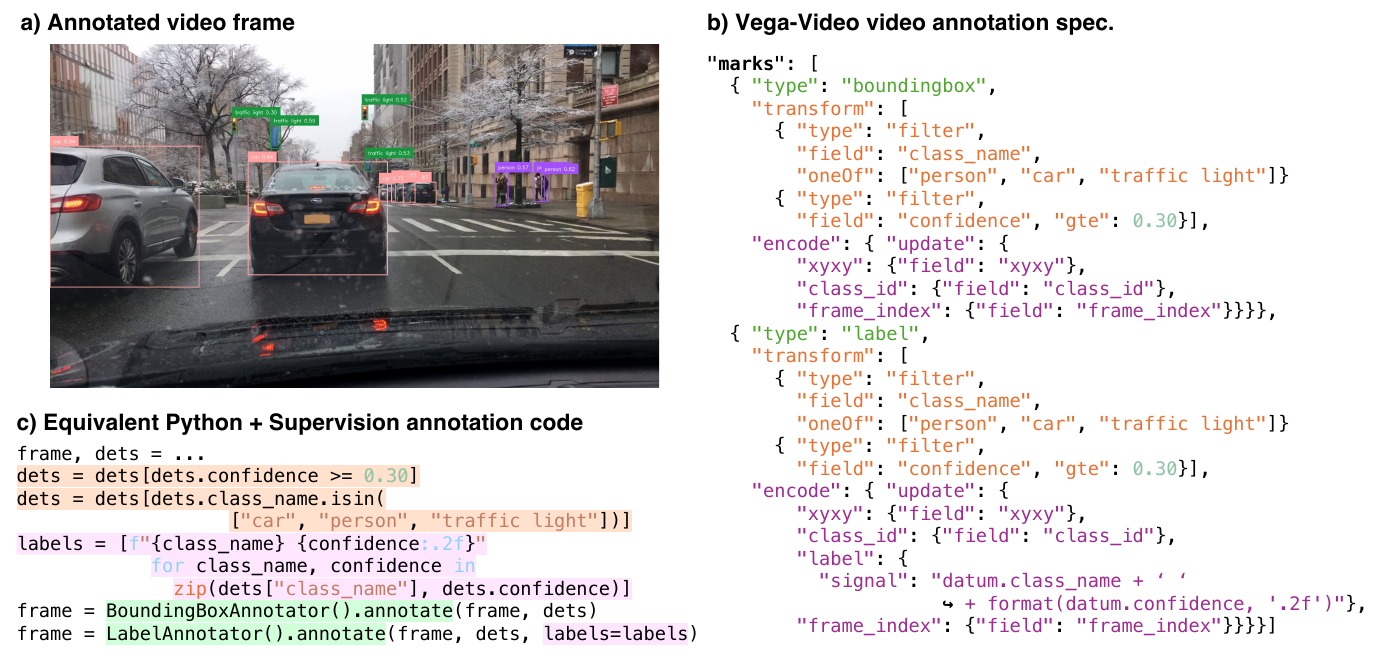}
    \caption{Frame annotation specification using our proposed grammar (b), as well as the equivalent Python + Supervision code (c), for drawing bounding boxes and a custom-formatted text label on objects in a video frame (a).}
    \label{fig:annotation-code}
\end{figure*}

To support video annotation, we conceptually build on datasets and marks.
A video is a 3-dimensional space with axes for x, y, and time, so a natural extension is to overlay plots atop a video's coordinate space.
In essence, our prior syncing approach directly applies by rendering on top of a video player.
This supports simple text overlays and rendering overlaid plots on the video, but is otherwise ill-suited for vision-based workflows.
Vision workflows have specific, known annotation approaches which, while similar to Vega's marks, are far from idiomatic.

Thus, we introduce a set of domain-specific video annotation marks for vision.
We use the Supervision~\cite{Roboflow_Supervision} project's Detections and Annotators model as inspiration.
Supervision models an individual detection as an \texttt{xyxy} field (object bounds), a mask (a bitmap of which pixels are part of the detection), confidence (from 0.0 to 1.0), class ID, class name (e.g., ``person''), and a tracker ID for cross-frame continuity.
Each frame corresponds to a set of these detections.
Further, Supervision has a collection of Annotators that can render different varieties of detections, such as BoundingBoxes and Labels.
Note that these are designed for composition, with the appropriate parameters for high-level behaviors, such as where, relative to the box, to place a text label.
Code using Supervision to annotate a frame is shown in \Cref{fig:annotation-code} (c).

In our specification grammar, we use a per-player \texttt{"marks"} attribute for video annotations.
A dataset of vision data can be mapped to the `Detections' data model, with a frame index to store the temporal position, and then marks corresponding to each annotation type render the corresponding overlays.
We show an annotated frame, and the corresponding Supervision and Vega-Video specification in \Cref{fig:annotation-code}.
This approach allows filtering of annotations, such as adding a minimum-confidence slider or showing only annotations that intersect a selected region; both use the existing Vega transformation grammar.
From an implementation perspective, annotation is a straightforward linked view problem: updating annotations is no more complex than updating a scatter plot.

\subsection{Video Transformation}\label{sec:transform-grammar}

The third pillar of our grammar describes \emph{video transformation}, enabling interactions to update video \textit{content}, not just the video playback state.
While syncing and annotating are established techniques, video transformation paradigms are, to the best of our knowledge, novel.
Therefore, we first propose some interactions to introduce and motivate the concept.

\begin{figure*}
    \centering
    \includegraphics[width=\linewidth]{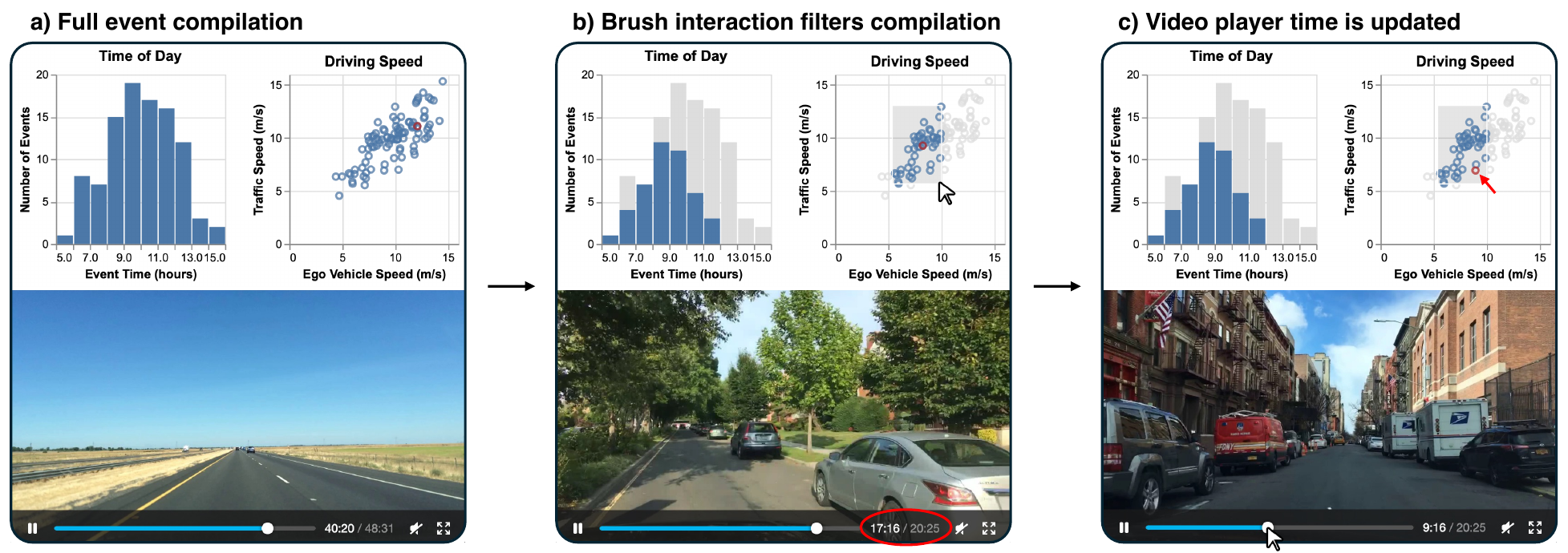}
    \caption{Brushing and linking with a video compilation. A brush selection (b) updates both a linked plot and a linked video compilation. A marker on the scatter plot follows the currently displayed event in the compilation (c).}
    \label{fig:brush_and_link_video}
\end{figure*}

In \Cref{fig:brush_and_link_video}, we detail brushing and linking~\cite{becker1987brushing} with a video compilation.
A dataset of driving events is spliced together into a long video, while per-event data is plotted.
The scatter plot renders the event currently being displayed by the video player in a highlight color.
Similarly, clicking on an event in the scatter plot jumps to the corresponding point in the video.
Both of these are syncing tasks.
Users can brush to select regions on these plots, and this updates both the other plots and the video compilation; the more events excluded, the shorter the video becomes.

In addition to filtering, another applicable operation on video compilations is sorting.
Sorting, or re-sorting, a compilation allows users to see how the ordered-by attribute affects the video content.
For example, sorting by vehicle speed allows users to quickly jump to different speed percentiles within the dataset.
This builds on a user's existing understanding of video playback: users intuitively find a specific part of a long video by conducting an informal binary search, and they can also promptly skim unfamiliar videos to form a general impression.

Video transformations are not limited to discrete segments; continuous transformations are also possible.
For example, given long-form drone flight recordings and telemetry data, it may be helpful to have a minimum altitude slider to filter a video.
As the cutoff value increases, the video becomes shorter because fewer frames fall within the range.

\begin{figure}
    \centering
    \includegraphics[width=\linewidth]{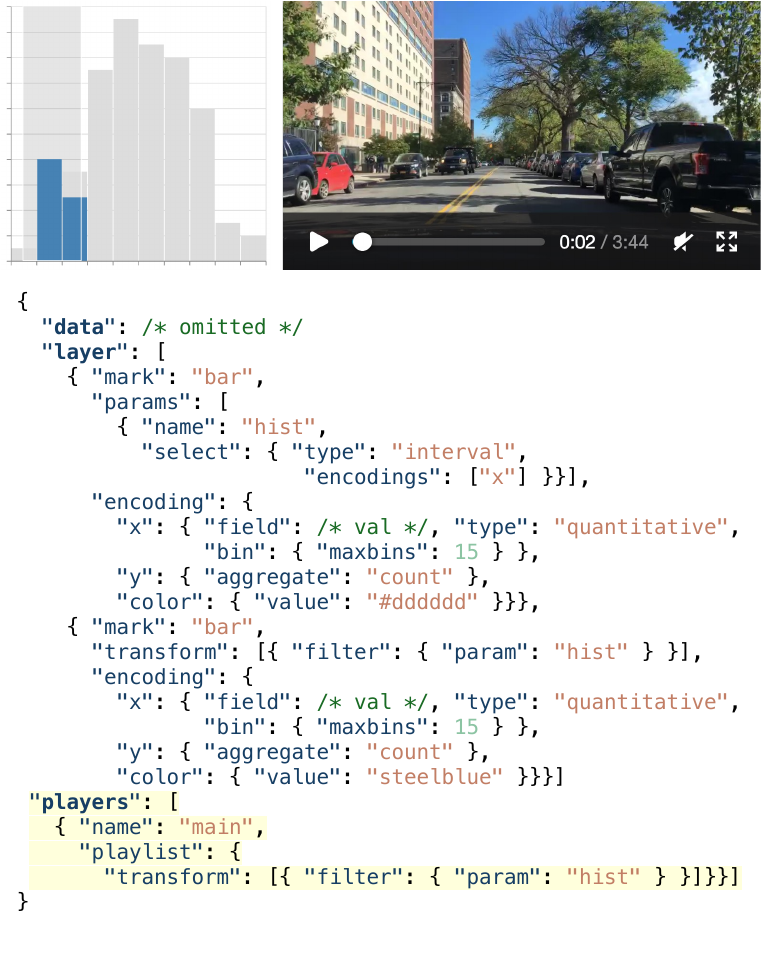}
    \caption{A Vega-Lite specification for brushing a histogram and linking a video compilation. Only a \texttt{"players"} object needs to be added, using existing data transforms.}
    \label{fig:bandl-code}
\end{figure}

We extend our grammar to support these types of interactions.
Rather than providing a source to a video player, we provide a \texttt{"playlist"} object which applies transformations.
We support a ``clip'' transform, which extracts a segment, as well as the existing ``sort'' operator.
At the top level, each video and clip is spliced together.
This allows us to easily convey what the actively playing segment is in the specification and why we refer to the transform's output as a playlist.
A minimal example of filtering a compilation is shown in \Cref{fig:bandl-code}.
Representing brushing and linking behavior in Vega is verbose, so we demonstrate this using the Vega-Lite version of our grammar.
Additional signals for playlist state, such as the current segment and the \texttt{time} and \texttt{itime} \textit{within} a segment, ensure common patterns do not require complex in-specification mappings.
Additionally, we find that one-dimensional erosion and dilation video transforms are particularly effective for removing segments with only a few frames and for adding context around segments.

\subsection{Cursor Consistency}
Once a video is updated from a previous playlist assembly, we need to set an initial playback time.
In \Cref{fig:brush_and_link_video}, as the video is filtered between (a) and (b), the video's time remains at roughly the same proportion throughout the video, even though the video becomes much shorter.
This is more appropriate than restarting the video from the beginning every time.
Additionally, due to segment caching, maintaining this temporal locality can further reduce the latency of rendering a transformed video.
We refer to this as \textit{cursor consistency}.

For some visualizations, such as filtering whole segments in \Cref{fig:brush_and_link_video,fig:bandl-code}, cursor consistency policies are straightforward.
If the currently displayed segment remains in the compilation, begin playback on the same segment at the same time.
If the currently displayed segment is filtered out of the compilation, begin at the first frame of a nearby segment, preferably the first segment after the current one that was not filtered out.
Similarly, for reordering segments, playback should always resume in the same segment and at the same relative time within it.
Both of these cursor consistency policies provide visualization consistency for users and temporal locality for segment caches.
Future user studies may be needed to verify the former, but the naive approach is clearly insufficient.

To specify cursor consistency behavior, we introduce a \texttt{"cursor"} field to a video player to set the target time after a transformation.
If the current segment remains in the playlist, \texttt{onKeep} sets the time; if the current segment is removed, \texttt{onRemove} sets the time instead:

\vspace{2mm}
{
    \centering
    \includegraphics[width=\linewidth]{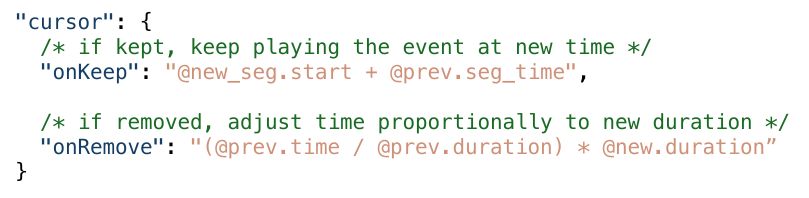}
}
\vspace{-2mm}

Since this is a transition behavior, both the previous and the new state are available.
This calculation has access to the old video player state via \texttt{@prev}, which contains time information, and partial \texttt{@new} video, which contains only a duration (i.e., no time set).
Additionally, \texttt{@prev\_seg} and, inside \texttt{onKeep}, \texttt{@new\_seg} provide segment bounds, with each supporting lead/lag window functions to reference nearby segments in their respective orderings.
This cursor consistency grammar reduces development viscosity~\cite{10.1145/1294211.1294256}.
Common behaviors are expressed in a single line and easily iterated on, while the full expressiveness is available for unforeseen policies.
Without native cursor consistency handling, this behavior is prohibitively complex to implement.

\section{Synchronizing Visualizations With Videos}\label{sec:syncing}

We now turn from our declarative grammar to the mechanics of ensuring visualizations and videos remain responsive and in sync.
Doing so requires reconciling two different systems:

\begin{itemize}
  \item A \emph{reactive dataflow graph} that assumes instantaneous, deterministic propagation of state with sub-millisecond updates.
  \item A \emph{stateful video subsystem} which may take hundreds of milliseconds, or even seconds, to respond due to seeking, decoding, and network latency.
\end{itemize}

Interactions may emerge from either of these systems, and the state updates must be reflected across both.
These systems maintain multiple values of state, particularly values of time, which need to be unified while preserving the semantics of a declarative specification.

\subsection{Unifying \& Reconciling State}\label{sec:time}

Externally, we maintain two time signals, one for intent, \texttt{@player.itime}, and one for the currently displayed frame, \texttt{@player.time}.
However, internally, there are multiple discrete time values for each video player.
In the dataflow graph, each signal is split into nodes, consolidating all reads into one and all writes into the other.
These map to lower-level operations, such as \texttt{@player.time} resolving to \texttt{@player.seek\_to} on a write and \texttt{@player.current\_time} on a read.
These lower-level signals are used by a video bridge controller, which connects the dataflow graph to a video player.

\begin{figure*}
    \centering
    \includegraphics[width=1.0\linewidth]{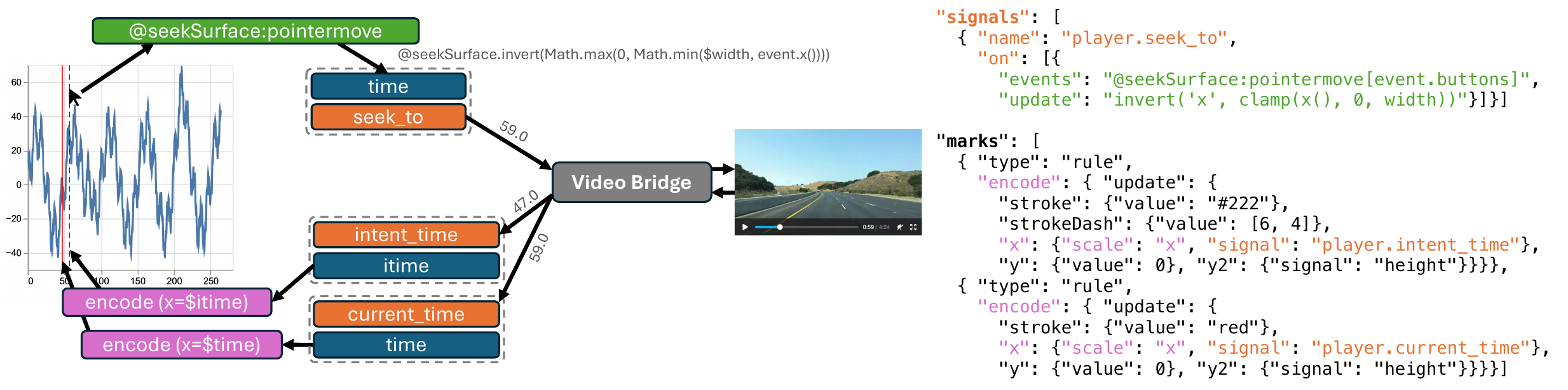}
    \caption{Conventional and video visualization synchronization architecture. Rewritten specification from \Cref{fig:syncing-code} shown.}
    \label{fig:syncing-arch}
\end{figure*}

\Cref{fig:syncing-arch} shows our synchronization architecture applied to the example in \Cref{fig:syncing-code}, including a specification showing the underlying signals.
An interaction on the seek surface triggers a write to \texttt{@player.seek\_to}, which is relayed to the video player through the bridge.
The video player then maps the seek request time to an in-video timestamp and updates the timeline progress bar location (which, conceptually, displays \texttt{@player.itime}).
The bridge then takes the updated in-video timestamp and updates \texttt{@player.intent\_time}, which the specification sees as \texttt{@player.itime}, triggering a recomputation on the dataflow graph and rendering an updated visualization.
This occurs in microseconds since no video data processing occurs in this pipeline.
The player still reflects the prior frame, but begins loading the newly requested frame in the background.
Only once this updated frame is ready is an updated \texttt{@player.current\_time} written, updating any graphics depending on \texttt{@player.time}.

This approach serves several purposes.
First, updating a graph with proxy nodes maintains the benefits of a dataflow graph while interfacing with an external system.
The bridge and video player can act as an extension of the dataflow graph, propagating necessary state changes while omitting synchronization for events/signals that are constant or unused.
Second, by splitting signals, we can introduce delayed updates without blocking the rendering pipeline.
Time updates are enqueued in the player, eventually triggering rendered visualization updates while maintaining semantically correct behavior.
Third, this precludes cycles while still allowing interactions to originate from both the conventional visualization and the video player.
For example, using an individual signal without splitting would create race conditions in updates and potentially create feedback loops.
Instead, this forces a Vega Interaction $\rightarrow$ Video Player Interaction $\rightarrow$ \{Video Player Presentation,  Vega Presentation\} ordering on data flow.
Fourth, and finally, this allows for reconciling differences in the domains of values.
An interaction on a seek surface, for example, will almost never produce the actual presentation time of a frame from a continuous range.
Or, an interaction may go beyond the surface and yield times beyond the span of the video, such as seeking to negative time.
These are troublesome as they violate the semantics of \texttt{@player.itime}, which should always refer to the frame next intended to be displayed.
Our synchronization architecture resolves these issues by allowing arbitrary updates to signals such as \texttt{@player.itime}, while ensuring that any dependent visualizations use the correct frame-exact values from the video player.

In addition to the two time signals, \texttt{@player.playing} uses the same split-signal approach, as do all writable playlist signals, including both presented time and intended time variants.
Non-writable signals, such as those at the tail of \Cref{tab:vega-signals}, have no semantically correct behavior for updates from a specification, so we reject any specification that attempts to write to these signals at compile time.

\subsection{Optimizing Scrubbing}\label{sec:scrubbing}

\begin{figure}
    \centering
    \includegraphics[width=0.85\linewidth]{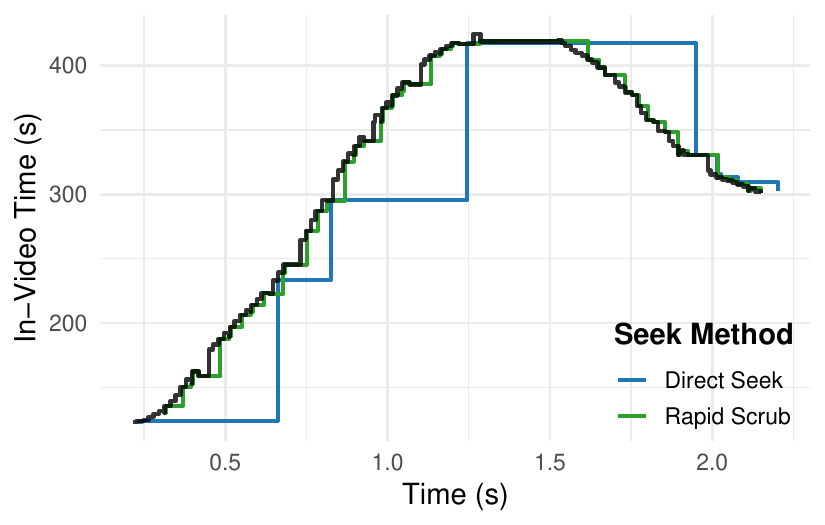}
    \caption{Seeking methods during scrubbing. The black line represents the intended time captured from a real event. Direct seeking has a substantial lag, whereas our Rapid Scrubbing method remains responsive.}
    \label{fig:seeking-methods-example}
\end{figure}

Seeking is the interaction type most likely to violate interactive latency constraints.
Playing, pausing, annotating, etc., all occur at sub-millisecond to single-millisecond time scales.
By contrast, seeking a video to a new position may require fetching new video segments over a network or from storage and decoding them, often taking hundreds of milliseconds, and potentially multiple seconds.
The most latency-sensitive cases are continuous interactions that rapidly seek the video, referred to as \textit{scrubbing} a video.
Consider \Cref{fig:seeking-methods-example}, where the intent time is shown in black and the presented time is shown in blue, and there is a substantial lag when using direct seeking.
We build on our prior synchronization architecture to identify such continuous seek interactions and use efficient keyframe-aware seeking strategies to ensure responsive visualizations.
Our objective is to ensure that seeking interactions remain responsive: minimizing the difference between the presented time and the intended time, i.e., \texttt{@player.time} and \texttt{@player.itime}.

\subsubsection{Video Encoding}

Modern video codecs encode frames differentially, allowing similar frames to deduplicate content.
To ensure seeking remains possible, and to restrict memory use, these are limited to short consecutive spans called Groups of Pictures (GOPs).
Each GOP begins with a keyframe, a frame with no dependencies that can be independently decoded.
When scrubbing, video players quickly fetch and decode GOPs, then abandon the task and try to display another.
However, since keyframes can be quickly decoded, seeking to keyframes is considerably faster.

\subsubsection{Detecting Continuous Seeks}

To apply appropriate seeking strategies, we classify the origin of writes to a video player's time at compile time.
Conceptually, we walk backwards up the dataflow graph, classifying an interaction as continuous only if it consists exclusively of known continuous operations.
For example, the specification shown in \Cref{fig:syncing-code} and \Cref{fig:syncing-arch} seeks to the following time on a \texttt{pointermove} event from this dataflow expression:

\vspace{2mm}
$\texttt{@seekSurface.invert}\!\left(\max\!\left(0,\; \min\!\left(\text{\$width},\; \text{event.x}\right)\right)\right)$
\vspace{2mm}

This is classified as a continuous seek since:

\begin{enumerate}
    \item The event, \texttt{pointermove}, is temporally continuous.
    \item The source $\text{event.x}$ is mathematically continuous.
    \item Composed invert, min, and max functions are continuous.
\end{enumerate}

Alternatively, an example of a discrete seek would be clicking on a scatter mark to jump to the corresponding event in a compilation.
This is discrete since a \texttt{click} event is temporally discrete, and because the time source, a static dataset value, is discrete.
In this case, seeking to the exact frame is required.

We incorporate this test as part of our specification rewrite previously detailed in \Cref{sec:time}.
A discrete or continuous tag is included with writes to \texttt{@player.seek\_to} and passed to the video bridge.

\subsubsection{Seeking Strategy}

Given a seek time and an interaction class, the video bridge controller executes one of two seeking strategies.
For discrete seeks, we use a conventional seek by jumping to the corresponding time in the video.
For continuous scrubbing, we adapt a keyframe-aware strategy.
Intuitively, we allow the video's time to ``snap'' to nearby keyframes that trail the user's intended position, frequently updating during rapid scrubbing, but settling into an exact location once the interaction settles.

\begin{figure}
    \centering
    \includegraphics[width=0.9\linewidth]{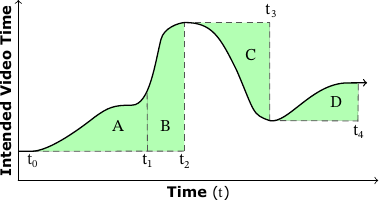}
    \caption{Keyframe search region during scrubbing. The solid line shows a user scrubbing a video's position. Presented frames must trail the user's intent, not jump ahead. At time $t_1$ seeking must occur in area $A$, at time $t_2$ in either $A$ or $B$, at time $t_3$ in area $C$, or at time $t_4$ in area $D$.}
    \label{fig:seeking-search-region}
\end{figure}

While scrubbing, the decoder is thrashed by updates, delaying the presentation of video frames for an unknown period.
However, this limits the updated frames to the region scrubbed over, essentially randomly sampling frames from this region to display.
\Cref{fig:seeking-search-region} shows this region, from the start of an interaction at $t_0$ to $t_2$ (past $t_2$, the entire range could occur).
Rather than arbitrarily sampling frames from this range through emergent thrashing behavior, we snap to keyframes that \textit{could have been} selected by this behavior.
By restricting seeks to keyframes, the expected time to decode a frame for presentation is significantly reduced, as is the update frequency, \textit{both} of which help mitigate thrashing.
This has two benefits: the first is higher-frequency, lower-error frame rendering while scrubbing.
The second benefit is a second-order effect: higher-frequency seeking loads new GOPs faster, including the ending GOP, where the interaction settles to an exact time.
By starting to load the keyframe of the ending GOP sooner, video data is fetched and a keyframe decoded earlier, allowing subsequent frames to decode faster.

During scrubbing, we calculate the keyframe search region based on the current time, the intended time, the keyframe times, and the start time of the current movement.
Let:

\begin{itemize}[leftmargin=0.75cm]
    \item $K=\{k_0, \dots, k_{m-1}\}$ be the keyframe timestamps in the video;
    \item $p$ be the currently presented frame's time (\texttt{@player.time});
    \item $t$ be the currently intended video time (\texttt{@player.itime});
    \item $\text{dir}\in \{+1, -1\}$ be the direction of the current scrub;
    \item $t_s$ be the intended video time at which the scrub began or last changed direction.
\end{itemize}
We then define a keyframe interval $I(t)$ as,

$$
  I(t) =
  \begin{cases}
    \left[\max\{p,\,t_s\},\, t\right] & \text{if } \mathrm{dir} = +1 \quad\text{(scrubbing forward)} \\[4pt]
    \left[t,\, \min\{p,\,t_s\}\right] & \text{if } \mathrm{dir} = -1 \quad\text{(scrubbing backward)}
  \end{cases}
$$

The illustration in \Cref{fig:seeking-search-region} shows how this keyframe search region changes with movement following the start of an interaction at $t_0$, assuming no intermediate frames are rendered before $t$.
Given $I(t)$, if $K \cap I(t) \neq \varnothing$, we choose the snapped keyframe $k^*(t)$ as the closest keyframe to $t$:

$$
  k^\star(t) =
  \begin{cases}
    \displaystyle \max \{\,k_j \in K \mid k_j \in I(t)\,\} &
      \text{if } \mathrm{dir} = +1 \\[8pt]
    \displaystyle \min \{\,k_j \in K \mid k_j \in I(t)\,\} &
      \text{if } \mathrm{dir} = -1
  \end{cases}
$$

Otherwise, if $K \cap I(t) = \varnothing$, we fall back to a conventional exact seek to $t$.

We apply keyframe snapping until we hit a time threshold with no activity, then we trigger an exact seek.
Scrubbing video is a commonplace occurrence in multimedia systems; however, interactive video data visualizations have far more interaction, particularly seeking and scrubbing, than conventional long-form video streaming, necessitating improved responsiveness.
Additionally, incorrectly applying keyframe scrubbing to discrete interactions adds latency and displays incorrect frames temporarily, making it less applicable in conventional applications.
However, in our case, we can filter out such discrete interactions by analyzing the visualization data flow in advance.
Long-form video streaming uses other methods to improve seeking responsiveness, such as initially loading lower-quality video segments or displaying a low-resolution preview thumbnail over the playback progress bar's cursor; both approaches could apply to our use cases as well.

\subsection{Transforming Video Source Selections}\label{sec:virtvideo}

\begin{figure*}
    \centering
    \includegraphics[width=\linewidth]{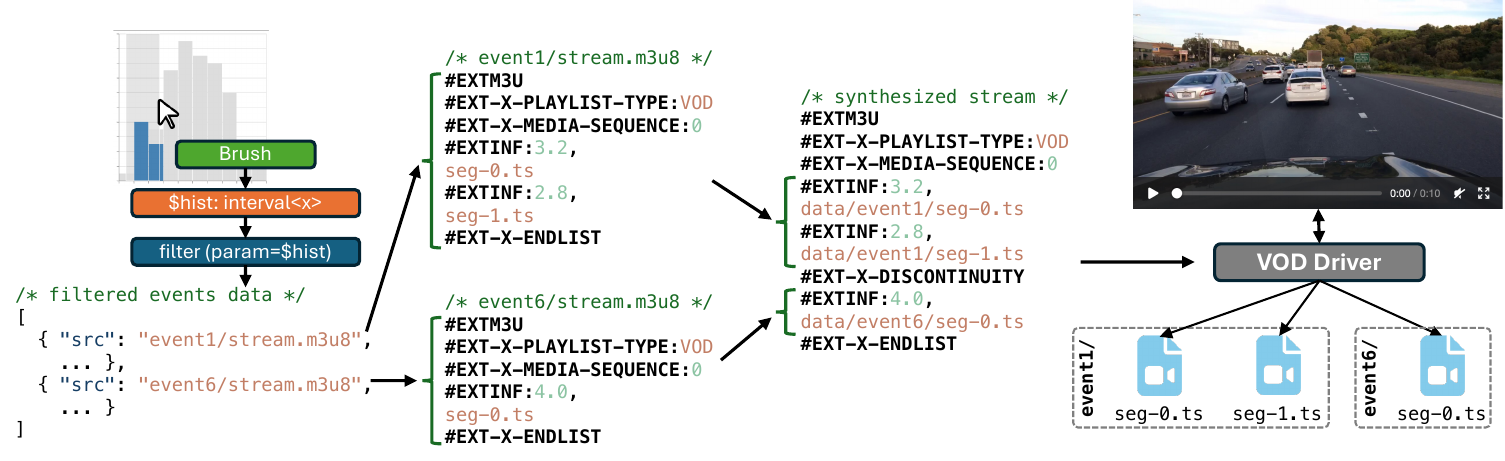}
    \caption{Video transformation through VOD manifest rewrites; showing the visualization in \Cref{fig:bandl-code}. A brush interaction filters out all but two data tuples. These tuples link to their video manifests, 6 and 4 seconds long, respectively. These event videos are spliced together to create a synthesized stream with a transformed 10-second video.}
    \label{fig:vod-updates}
\end{figure*}

Video data visualizations need to transform video \textit{content} in real time, not just synchronize with video playback or overlay annotations.
Doing so requires performing lightweight video editing, an area that poses significant performance challenges, especially in a web browser fetching remote video resources.
Rather than directly manipulating pixels or frames, we build on Video on Demand (VOD) protocols to implement the transformations detailed in \Cref{sec:transform-grammar}.

We use HTTP Live Streaming (HLS)~\cite{rfc8216} in our examples, one of the two major VOD standards.
Other VOD protocols, such as DASH~\cite{ISO23009}, are equally applicable.
VOD declaratively describes a video stream via a manifest file that contains links to short media segments and their durations.
Such manifest files are tiny compared to the underlying media; hours of video can be represented in kilobytes.
Crucially, the principles governing Video on Demand are complementary to our use case:

\begin{itemize}[leftmargin=0.75cm]
    \item Segments start on keyframes and can be fetched independently.
    \item Videos can be split into VOD segments quickly, without transcoding, or even repackaged on request.
    \item Video segments can be cached, both by clients and by content distribution networks, even across manifest changes.
    \item Manifest files are declarative and minimally expressive, enabling efficient hardware-accelerated playback.
    \item Manifest files are designed to be updated in use, since that is how live streaming is implemented.
    \item VOD standards can splice many differently-encoded streams together, ``stream discontinuities'' in HLS parlance, as advertisement injection builds on this.
\end{itemize}

We implement video transformations as transformations on VOD manifest files.
Each span of a video is represented as an in-memory collection of VOD segments, each with a segment URL and duration, upon which we apply clipping and splicing operators, consolidating these into a synthesized stream manifest.
\Cref{fig:vod-updates} provides an example of these rewrites for the brushing and linking example from \Cref{fig:bandl-code}.
A user interaction updates the filtering interval signal \texttt{hist}, which propagates through the filter transformation, leaving only the events that fit the condition.
Each of these events contains a VOD stream for the event video, which is spliced into a synthesized VOD stream manifest file.
This manifest is updated within the HLS VOD Driver, which updates the video player and issues requests to fetch segments as needed.

\section{Tradeoffs and Limitations}

\subsection{Video in Web Browsers}

Given the use-inspired nature of our research, we have made pragmatic decisions when building our extension due to external technical constraints.
Since video data visualizations are inherently interactive, use is constrained to web browsers.
Video has special treatment in web browsers.
While combining an HTML \texttt{<canvas>} element with general-purpose computing via JavaScript or WebAssembly makes anything theoretically possible, performance and energy efficiency ensure that serious applications delegate video to a \texttt{<video>} element and its associated APIs, particularly HTML5's \texttt{MediaSource}, which is explicitly designed to provide hardware-accelerated decoders to VOD clients.
As a result, our implementation \textit{requires} videos to be served over VOD protocols to support video transformation, although syncing-only applications can use static media files.

Due to our use of \texttt{<video>} elements, we are unable to perform frame-exact playback synchronization \textit{between} video players.
This is a fundamental operation, especially in multi-camera applications, and is cleanly expressed in our grammar.
Additionally, raster-based annotations, such as overlaying heatmaps or object masks, depend on joined video streams.
However, these are impossible to perform in a web browser without sacrificing either correctness or efficiency under the existing APIs.
As a workaround, a control loop can provide best-effort synchronization.
Frame-exact approaches require manual implementation of video decoding and compositing.
This carries a significant performance penalty.
Alternatively, one can compose multiple videos into a single stream and split them on the frontend.
This requires significant video preprocessing.
The emerging WebCodecs API could lift these constraints by providing direct decoder access, but at the cost of forgoing the highly optimized VOD playback pipeline and potentially competing with Vega's own dataflow and rendering for resources.

\subsection{Video Transformation}

Our video transformation approach is limited to operations expressible as manifest edits over VOD streams, such as filtering, reordering, and clipping at segment boundaries.
Richer transformations, including raster annotation compositing and multi-video composition, are constrained by the capabilities of browser-based video playback.
These limitations can be alleviated by offloading some video transformation to a server, analogous to VegaFusion/VegaPlus for data transforms.
On-demand segment transcoding is already standard practice in Video CDNs, where lazy transcoding and remuxing avoid pre-encoding every format variant~\cite{mux_justintime}.
Vidformer~\cite{vidformer} extends this pattern beyond transcoding to render transformed and annotated segments for video data visualization.
From the browser's perspective, nothing changes: it receives ordinary VOD segments and uses hardware-accelerated playback as before.
The tradeoff is increased latency, on the order of hundreds of milliseconds per segment~\cite{vidformer}.

\subsection{Correctness in Frame Times}

Floating-point time is the most straightforward way to link conventional visualizations and video playback, but it poses correctness challenges.
Framerates from NTSC regions are factors of 3, and in PAL regions, factors of 5; neither of these creates fractions of time exactly representable in floating point.
Combining this with floating-point arithmetic's lack of associativity, using floating-point numbers for selecting discrete frames poses inherent challenges for ensuring frame-exact synchronization.
As a remedy, in addition to floating-point signals like \texttt{@player.time} and its many variants, we include corresponding signals using frame indexes.
This partially addresses these challenges, but at the expense of simplicity, and it remains impractical for variable frame rate (VFR) videos.

\subsection{Syncing with Audio}

When wireless audio devices are connected, browsers delay video frames to compensate for audio transmission latency.
While this has minimal impact on regular long-form video playback, applications in video data visualization may be disproportionately affected.
This creates an unusual situation where the responsiveness of an interactive visualization depends on the presence and model type of a wireless audio device.
We measured 136 ms of additional latency with Apple AirPods Pro 2 and 147 ms with Bose QC45s.
Newer audio codecs reduce this latency, and purpose-built low-latency audio codecs reduce it significantly, but this is only a latency reduction, potentially at the expense of audio quality.

\section{Evaluation}

\subsection{Datasets}

For syncing evaluation, we use the Blender Tears of Steel (ToS) reference video.
It is $1280\times720$, 24 FPS, and $734$ seconds long, and is 56 MB encoded on disk.
It was encoded with \texttt{x264} using default FFmpeg flags, averaging one keyframe every 4.37 seconds.
It is stored in a \texttt{.mp4} file with a moov atom at the front (i.e., FFmpeg's ``\texttt{faststart}'' flag).
Audio is encoded in stereo AAC at 130 kbps.

For video collections, we use the Berkeley Deep Drive 100k dataset~\cite{yu2020bdd100k}.
These are $1280\times720$ and 30 FPS, encoded with the same configuration as ToS, but split into 2-second HLS segments instead.
We use actual BDD-provided metadata and telemetry in these visualizations.

\subsubsection*{Environment}

Our evaluation was conducted on a system with an Apple M1 processor, 16GB of RAM, running macOS 26.1.
We used Vega \& Vega-Lite 6.

We stored all videos in AWS S3 and accessed them via the AWS CloudFront CDN over a residential internet connection with an average round-trip latency of 30.7ms, typical for a CDN.

\subsection{Results}

The trace in \Cref{fig:seeking-methods-example} shows a visible lag during scrubbing when using a direct seeking method (blue).
Our rapid scrubbing method is shown in green and, qualitatively, tracks the intended time more closely.
In this example trace, direct seeking has an average deviation of 45.4 seconds, while rapid scrubbing has an average deviation of 4.34 seconds.

\begin{table}[htbp]
    \centering
    \caption{Scrubbing method comparison across browsers. Chrome lacks native \texttt{fastSeek}, so we omit this baseline.}\label{tab:scrubbing-dat}
    \begin{tabular}{llcc}
        \toprule
        \textbf{Browser} & \textbf{Method} & \textbf{Avg. Deviation (s)} & \textbf{Avg. Updates/s} \\
        \midrule
        Firefox & Direct          & 103.7 & 2.21 \\
        Firefox & Keyframe        & 24.5  & 7.40 \\
        Firefox & Rapid (ours)   & 31.8  & 7.99 \\
        \midrule
        Chrome  & Direct          & 44.3  & 5.03 \\
        Chrome  & Keyframe        & n/a  & n/a \\
        Chrome  & Rapid (ours)   & 24.7  & 9.50 \\
        \midrule
        Safari  & Direct          & 46.4  & 7.48 \\
        Safari  & Keyframe        & 26.6  & 9.68 \\
        Safari  & Rapid (ours)   & 42.0  & 7.85 \\
        \bottomrule
    \end{tabular}
\end{table}

\begin{figure}
    \centering
    \includegraphics[width=\linewidth]{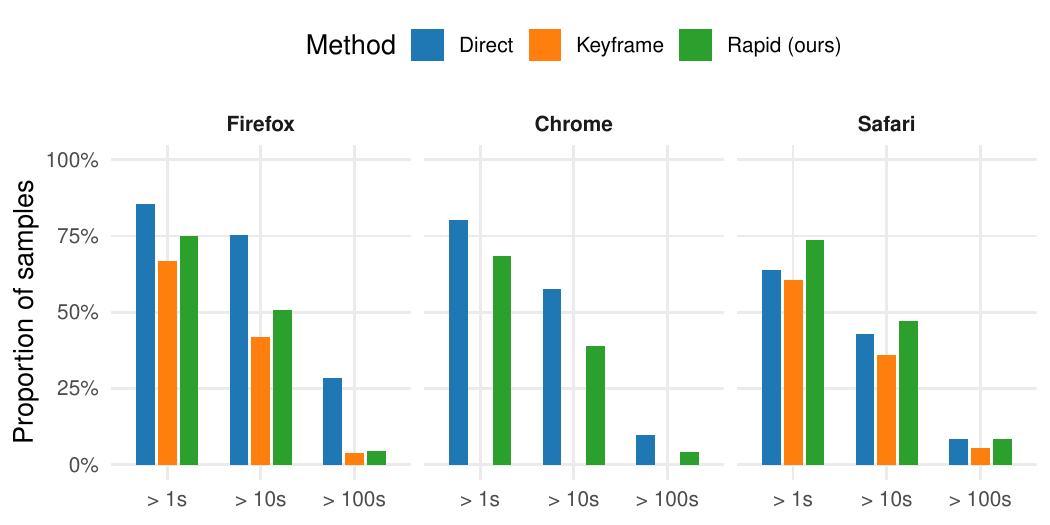}
    \caption{Scrubbing time deviation proportions. Nearest-keyframe and our rapid scrub method primarily reduce high-deviation events.}
    \label{fig:seeking-props}
\end{figure}

We captured 15 distinct scrubbing traces using a visualization similar to that in \Cref{fig:syncing-code}.
Five were towards specific points, five were undirected searches across the video, and five were sense-making exploration, scrubbing around the entire video.
We evaluated the visualization in the Firefox, Chrome, and Safari web browsers.
Results are shown in \Cref{tab:scrubbing-dat}, showing that both nearest-keyframe baselines and our rapid scrub methods have much higher update rates, resulting in lower time deviation during scrubbing than with direct seeking.
Notably, direct seeking averaged one update every 452ms, already exceeding even the most lenient interactivity latency thresholds.
In \Cref{fig:seeking-props}, we show the distribution of time deviation, confirming that a higher frame update rate reduces high-deviation events.
Safari, which uses a more aggressive seeking approach in our baselines, exhibited higher deviation.
Across all methods and browsers, the Vega side of the visualization rendered in ${\sim}2.3\text{--}3.6$ms on average, with p99 times at ${\sim}7\text{--}10$ms.

\begin{figure}
    \centering
    \includegraphics[width=\linewidth]{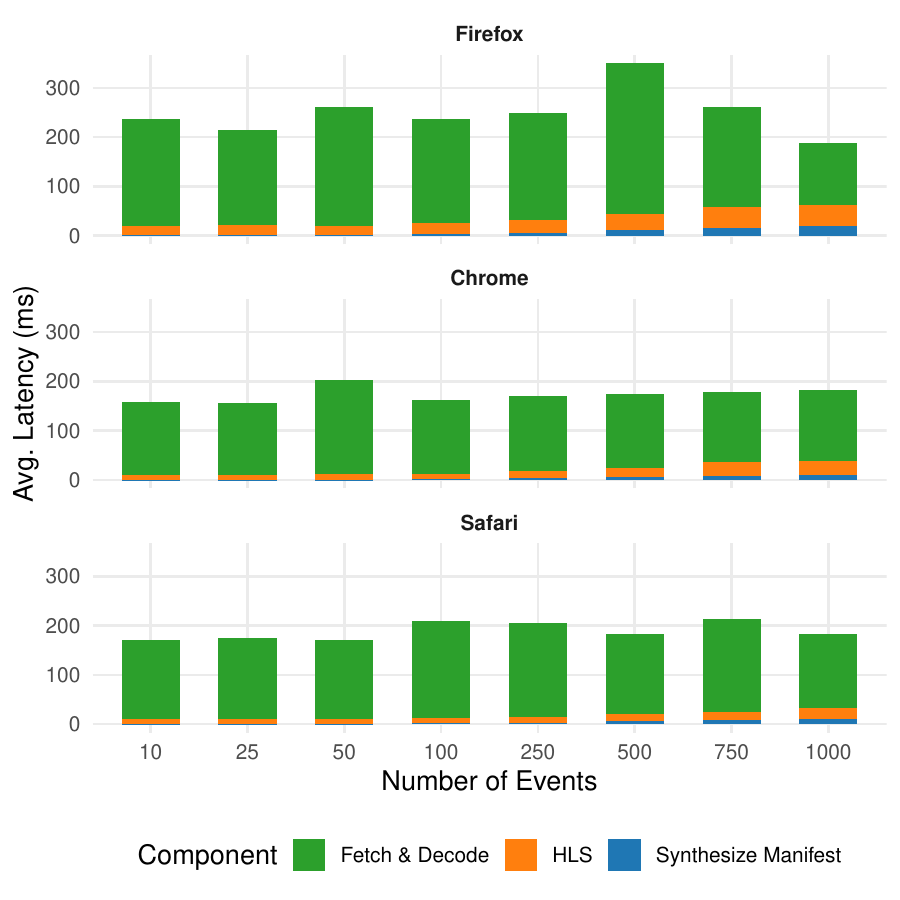}
    \caption{Time to transform a video after a filter event.}
    \label{fig:transform-browsers}
\end{figure}

We evaluated real-time video transformation using brushing and linking by implementing a visualization similar to \Cref{fig:brush_and_link_video}.
We evaluated the time to update the video player when selecting $n$ events, up to 1000, which results in a video over 10 hours.
Results are shown in \Cref{fig:transform-browsers}, averaged across 8 trials per configuration.
We measured an average of 189ms with 1000 events.
Varying the number of events made minimal difference, with only marginal additional time to synthesize the manifest and for the HLS driver to update with many events.

\section{Conclusion}

In this paper, we identify three classes of video visualizations: syncing, annotating, and video transformation.
We then integrate each of these into a Grammar of Graphics-based language, enabling combinations among these three classes and with existing classical data visualization methods.
We show that specification-level rewrites enable a semantically unified notion of time that masks the underlying synchronization delay.
Further, we show how declarative specifications can be leveraged to implement the optimizations necessary to achieve interactive latencies.
Specifically, rapid scrubbing reduces time deviation by $4\times$ and increases updates per second by $3\times$, while real-time video transformation completes on the order of ${\sim}175$ms.
We release our implementation of this work as open source, enabling expressive, interactive, and portable video data visualizations in practice\footnote{\url{https://github.com/ixlab/vega-video}}.
By making video a first-class data type in the Grammar of Graphics, we bring the composability, portability, and performance of modern data visualization to an increasingly critical modality.

\subsubsection*{Future Work}
This paper provides a foundation for constructing expressive and performant video/data visualizations, but further work is needed to determine which interaction patterns best serve real-world analysis tasks.
Video transformation and cursor consistency specifically deserve further attention due to their novelty.
Additionally, we anticipate that tuning VOD control algorithms for our use case will yield further performance improvements.

\acknowledgments{This material is based upon work supported by the National Science Foundation under Grant No.\ IIS-1910356 and the OAC-2118240 Imageomics Institute award.}

\bibliography{main}

\end{document}